\documentclass[journal, onecolumn,12pt, draftclsnofoot]{IEEEtran}

\usepackage{physics}
\usepackage{amsmath}
\usepackage{amsfonts}
\usepackage{mathtools}
\usepackage{amsfonts}
\usepackage{amssymb}
\usepackage{array}
\usepackage{makeidx}
\usepackage{csquotes}
\usepackage{graphicx}
\usepackage{cite}
\usepackage{xcolor}
\usepackage{subcaption}
\usepackage{graphicx}
\usepackage[acronym,toc,shortcuts]{glossaries}
\newacronym{TN}{TN}{Terrestrial Networks}
\newacronym{NTN}{NTN}{Non-Terrestrial Networks}
\newacronym{LEO}{LEO}{Low Earth Orbit}
\newacronym{ISAC}{ISAC}{Integrated Sensing and Communication}
\newacronym{HAPS}{HAPS}{High-Altitude Platform Stations}
\newacronym{LoS}{LoS}{Line-of-Sight}
\newacronym{UAV}{UAV}{Unmanned Aerial Vehicles}
\newacronym{3GPP}{3GPP}{3$^{\text{rd}}$ Generation Partnership Project}
\newacronym{CSI}{CSI}{Channel State Information}
\newacronym{6G}{6G}{Sixth Generation}
\newacronym{AoA}{AoA}{Angle-of-Arrival}
\newacronym{RIS}{RIS}{Reconfigurable Intelligent Surfaces}
\newacronym{MEO}{MEO}{Medium Earth Orbit}
\newacronym{GEO}{GEO}{Geostationary Earth Orbit}
\newacronym{BS}{BS}{Base Station}
\newacronym{IoT}{IoT}{Internet of Things}
\newacronym{AI}{AI}{Artificial Intelligence}
\newacronym{ML}{ML}{Machine Learning}
\newacronym{PHY}{PHY}{Physical Layer}
\newacronym{AoD}{AoD}{Angle-of-Departure}
\newacronym{ETSI}{ETSI}{European Telecommunications Standards Institute}
\newacronym{RAN}{RAN}{Radio Access Network}
\newacronym{NFC}{NFC}{Near-Field Communication}
\usepackage{multicol}
\usepackage{multirow}
\usepackage{booktabs}
\usepackage[table]{xcolor}
\definecolor{good}{RGB}{198,239,206}     
\definecolor{average}{RGB}{255,235,156}  
\definecolor{poor}{RGB}{255,199,206}     

\definecolor{headergray}{gray}{0.85}
\definecolor{lightgray}{gray}{0.95}
\definecolor{checkgreen}{RGB}{0,140,0}
\definecolor{crossred}{RGB}{180,0,0}

\newcommand{\cmark}{\textcolor{checkgreen}{\ding{51}}}
\newcommand{\xmark}{\textcolor{crossred}{\ding{55}}}
\usepackage{pifont}
\usepackage{array}

\begin{document}
%
\title{ISAC-Enabled Non-Terrestrial Networks for 6G: Design Principles, Standardization,  Performance Tradeoffs, and Use Cases}

%

\author{ Muhammad Ali Jamshed,~\IEEEmembership{Senior Member,~IEEE}, Rohit Singh,~\IEEEmembership{Member,~IEEE}, Malik Muhammad Saad, Aryan Kaushik,~\IEEEmembership{Member,~IEEE}, Wonjae Shin,~\IEEEmembership{Senior Member,~IEEE,} Miguel Dajer,~\IEEEmembership{Member,~IEEE}, Alain Mourad
\thanks{
M. A. Jamshed is with the College of Science and Engineering, University	of Glasgow, UK (e-mail: muhammadali.jamshed@glasgow.ac.uk).

R. Singh is with Dr B R Ambedkar National Institute of Technology, India (email: rohits@nitj.ac.in).

M. M. Saad is with the Department of Electrical Engineering and Computer Science,
Daegu Gyeongbuk Institute of Science and Technology (DGIST), South Korea (email: maliksaad@dgist.ac.kr).

A. Kaushik is with OUF Innovative, UK, RakFort, Ireland and IIITD, India (email: a.kaushik@ieee.org).

W. Shin is with the School of Electrical Engineering, Korea University, Seoul, South Korea (email: wjshin@korea.ac.kr). (Corresponding Author).

M. Dajer is with Futurewei Technologies, San Jose, CA 95131 USA (email: mdajer@futurewei.com).

A. Mourad is with InterDigital Europe Ltd., EC3A 3DH London, U.K (email: Alain.Mourad@interdigital.com).
}
}


\maketitle

\begin{abstract}
Non-Terrestrial Networks (NTN) have emerged as a key enabler to fully realize the vision of integrated, intelligent, and ubiquitous connectivity in 6G systems. However, several operational challenges, including severe Doppler effects, interference, and latency, hinder the seamless integration of NTN and Terrestrial Networks (TN). In this context, Integrated Sensing and Communication (ISAC), which unifies sensing and communication functionalities within a common framework, offers great potential to address these challenges while enabling new network capabilities. Due to its complementary functionalities, ISAC can play a pivotal role in enhancing NTN performance, although its practical adoption requires a fundamental rethinking of existing architectural and standardization frameworks. Motivated by this need, this article examines key aspects of ISAC-enabled NTN, including  architectural design principles, application scenarios, standardization challenges, and key performance tradeoffs. Finally, a representative case study is presented to illustrate major technical challenges and highlight promising future research directions for ISAC-enabled NTN.




\end{abstract}


\IEEEpeerreviewmaketitle


\section{Introduction} 

Every generation of wireless communications aims to outperform the previous one by delivering higher data rates, lower latency, and more reliable connectivity. However, achieving truly universal coverage remains a major challenge. In early 2025, approximately 2.9 billion people still lack mobile connectivity, particularly in remote and  sparsely populated areas where the deployment of terrestrial infrastructure is economically challenging \cite{jamshed2025tutorial}. As a potential solution, the \ac{3GPP} Release~17 stresses the ntegration \ac{TN} and \ac{NTN} to provide ubiquitous connectivity, thus laying the foundation for ubiquitous \ac{6G} systems to fulfill its designated requirements. However, several challenges limit this integration, including severe Doppler shifts, higher latency, interference, and spectrum management issues, especially in satellite-based NTN deployments  \cite{11372048}.

Among promising technological enablers, \ac{ISAC}, which unifies  communication and sensing functionalities within a shared framework, offers several complementary features that can address many of these challenges. 
ISAC has already demonstrated strong implementation potential by taking advantage of a large portion of the existing communication infrastructure. In parallel, \ac{3GPP} has initiated standardization efforts for \ac{ISAC} in Release~19 through technical reports and specifications that cover the relevant use cases and system requirements. Although \ac{ISAC} offers significant potential, such as quantifying integration and coordination gains, clock synchronization, phase offset mitigation, and improved resource management, several challenges still limit its full realization in practical \ac{6G} systems \cite{kaushik2024integrated}. 

Importantly, the synergy between NTN and ISAC is bidirectional. 
Although ISAC can help overcome key limitations of NTN, NTN itself can substantially enhance ISAC capabilities through wider coverage, improved \ac{LoS} availability, and richer spatial-temporal diversity. 
Due to their strong mutual complementarity, \ac{NTN} and \ac{ISAC} can be considered as two key enabling technologies for \ac{6G}. For example, \ac{ISAC} deployed solely within terrestrial infrastructure often suffers from blockage and limited sensing visibility due to obstacles, while integration with \ac{UAV} platforms has been shown to significantly improve sensing performance \cite{meng2023uav}. Similarly, in \ac{HAPS}-based \ac{NTN}, \ac{ISAC} can further improve latency, coverage, and spectral efficiency \cite{kanani2025haps}. In the context of \ac{LEO} satellite systems, recent studies have shown that ISAC can effectively address challenges, such as Doppler, energy efficiency, and security concerns
\cite{yin2024integrated}. 
 However, several important gaps remain, particularly in terms of standardization, architectural integration, and practical adoption across all NTN layers.

\begin{table*}
\centering
\caption{Positioning of This Work with Respect to Prior Literature}
\label{tab:comparison}
\renewcommand{\arraystretch}{1.3}
\rowcolors{2}{lightgray}{white}

\begin{tabular}{>{\columncolor{headergray}}p{3.8cm} c c c c c c}
\toprule
\rowcolor{headergray}
\textbf{Techniques} & ~\cite{meng2023uav} & ~\cite{kanani2025haps} & ~\cite{yin2024integrated} & ~\cite{11098638} & ~\cite{11017717} & \textbf{Our Work} \\
\midrule

\textbf{Specific NTN Use Case with ISAC}
& UAV & HAPS + UAV & LEO Satellites & UAV & UAV & \textbf{Complete NTN} \\

\textbf{Standardization Perspective}
& \xmark & \xmark & \xmark & \cmark & \cmark & \cmark \\

\textbf{NTN-ISAC System Architecture}
& UAV & HAPS + UAV & LEO Satellites & UAV & UAV & \textbf{Complete NTN} \\

\textbf{Multiplexing Techniques Comparison}
& \xmark & \xmark & Partial & \xmark & \xmark & \cmark \\

\textbf{Sensing Aided Communication Mechanism}
& \cmark & \xmark & \xmark & \xmark & \xmark & \cmark \\

\textbf{Performance tradeoffs Analysis}
& \cmark & \cmark & \cmark & \xmark & \xmark & \cmark \\

\bottomrule
\end{tabular}
\end{table*}








Previous discussion confirms how the combination of \ac{ISAC} and \ac{NTN} can provide a unified platform for sensing, monitoring, and data transmission \cite{meng2023uav, kanani2025haps, yin2024integrated}. Nevertheless, its implementation persists several associated challenges, e.g., accurate signal management, possibility of interference between communication and sensor, making this integration challenging, particularly in dynamic and resource-constrained environments such as spaceborne systems. 

Moreover, advancing this field requires answering a few critical questions about the use of \ac{ISAC} into \ac{NTN} in the context of 6G systems. \textbf{Q1:} How can ISAC fundamentally mitigate key challenges, such as Doppler, latency, and interference, beyond what communication-only NTN architectures can achieve? \textbf{Q2:} What architectural and waveform-level refinements are required to integrate ISAC into NTN? \textbf{Q3:} How does ISAC-enabled NTN enhance system-level performance and enable new/improved use cases compared to conventional NTN designs? \textbf{Q4:} What are the key challenges of standardization, implementation, and coexistence for ISAC-enabled NTN and emerging solutions?\footnote{ Compared to previous work, these questions are not limited to single NTN connectivity and also deal with the use of \ac{ISAC} in integrated \ac{TN} and \ac{NTN}.}    

In this article, we demonstrate how the use of \ac{ISAC} in \ac{NTN} can play a key role in fulfilling the promise of \ac{6G}. Specifically, we have provided a detailed comparison on how our work is positioned in relation to the relevant state-of-the art and is shown in Table \ref{tab:comparison}. Furthermore, the contributions of this article to the body of knowledge are summarized as follows:

\begin{itemize}
    \item This work provides a unified overview of ISAC and NTN, covering their fundamental operating paradigms and clarifying why ISAC is a key enabler to overcome NTN challenges, including Doppler effects, latency, interference, and sensing limitations.
    
    \item In addition, this work analyzes ISAC-enabled NTN architectures and representative use cases, highlighting how sensing–communication integration enhances coverage and coordination compared to conventional NTN designs.
    
    \item A detailed discussion of ongoing standardization activities for ISAC and NTN has been presented, with a focus on current 3GPP roadmaps, existing technical gaps, and open issues.
    
    \item Finally, a representative case study of ISAC-enabled NTN has been presented, illustrating sensing-assisted communication cooperation, followed by a discussion of critical technical challenges and promising future research directions.
\end{itemize}

\section{Role of ISAC for NTN}
\label{RIS_ISAC_NTN_sec}

\ac{ISAC} has recently emerged as a key enabler for \ac{6G} wireless systems, driven by the growing need to efficiently utilize limited spectrum resources while simultaneously supporting communication and environmental perception. By unifying radar-like sensing, data acquisition from multiple sensing sources, and data transmission within a shared radio framework, \ac{ISAC} enables wireless networks to evolve from passive communication infrastructures into active cyber-physical systems capable of observing, interpreting, and reacting to their surroundings.\footnote{It is important to note that \ac{ISAC} is not limited to the radar-like functionality provided by cellular network, but also encompasses the collection of information from diverse sensing sources and its exposure within the network as well as to higher-layer applications.} This paradigm shift is particularly significant for \ac{NTN}, where large propagation distances, high mobility, and dynamic channel conditions impose fundamental limitations on conventional communication-centric designs.

\subsection{ISAC Fundamentals \& Operating Paradigms}

In general, \ac{ISAC} implementations can be categorized into two major paradigms:
\begin{itemize}
    \item In the first paradigm, sensing and communication functionalities are realized using fully integrated hardware and waveform design, where a single radio front-end simultaneously transmits and receives signals for both purposes. This approach maximizes spectral efficiency and reduces hardware duplication; however, it also imposes stringent requirements on waveform design, transceiver architectures, and signal processing algorithms. 
    \item In the second paradigm, the sensing and communication subsystems operate independently while sharing the same frequency band, enabling greater design flexibility at the cost of increased interference management complexity. In such coexistence-based approaches, advanced interference suppression and coordination mechanisms are essential to ensure acceptable performance for both functions.
\end{itemize}

Although significant progress has been made in terrestrial \ac{ISAC} systems, most existing solutions rely heavily on favorable \ac{LoS} conditions and relatively stable channel environments. These assumptions do not always hold in \ac{NTN} scenarios, where links experience large Doppler shifts, long round-trip delays, atmospheric attenuation, and time-varying geometries. Consequently, directly applying terrestrial \ac{ISAC} designs to \ac{NTN} is insufficient, requiring a rethinking of \ac{ISAC} principles in the context of space–air–ground integrated networks.

\subsection{Why ISAC is Essential for NTN}

\ac{NTN} is inherently characterized by delays and frequently outdated \ac{CSI}, due to the long propagation distances between satellites, aerial platforms, and ground terminals. In such environments, conventional feedback-driven communication adaptation becomes ineffective, particularly under high-mobility conditions. \ac{ISAC} introduces an alternative paradigm in which sensing-derived environmental and mobility information complements, or even partially replaces, instantaneous channel measurements. By embedding sensing capabilities within \ac{NTN} platforms, the network can continuously estimate key environmental and geometric parameters such as platform positions, relative velocities, atmospheric conditions, and scattering characteristics. These parameters directly influence link quality and can be exploited to predict channel evolution, enabling proactive beam steering, Doppler pre-compensation, and adaptive resource allocation. As a result, \ac{ISAC} transforms \ac{NTN} from reactive communication systems into predictive and context-aware networks. This dual functionality is particularly attractive for satellite and aerial platforms, where payload, power, and spectrum resources are tightly constrained.
\subsection{Sensing-Aided Communication in NTN Environments}

Within an \ac{ISAC}-enabled \ac{NTN}, sensing and communication form a tightly coupled closed-loop system. Downlink and uplink transmissions inherently probe the surrounding environment, while resulting echoes and reflections are processed to extract sensing information such as delay, Doppler, and \ac{AoA}. These parameters, traditionally regarded as channel impairments, become valuable sources of situational awareness. Sensing outputs can be fed into network control and management functions to support communication decisions, including beam alignment, handover management, power control, and scheduling. For example, knowledge of satellite and user mobility patterns allows the network to anticipate link blockages and proactively adjust transmission strategies. In this way, sensing acts as an enabler of robust, reliable communication rather than as merely an auxiliary service.

\subsection{Enabling Technologies for ISAC in NTN}

Several emerging technologies play a crucial role in facilitating \ac{ISAC} operation in \ac{NTN} environments. \ac{RIS} can be deployed on satellites, aerial platforms, or terrestrial structures to enhance signal propagation, improve coverage in shadowed regions, and create favorable reflection paths for sensing \cite{10463684}. Advanced waveform designs that are resistant to large Doppler shifts and delay spreads are essential to maintain both sensing accuracy and communication reliability. Furthermore, edge intelligence and on-board processing capabilities enable local extraction of sensing features and real-time decision-making, thereby reducing backhaul overhead and end-to-end latency.



In summary, \ac{ISAC} plays a pivotal role in unlocking the full potential of \ac{NTN} by enabling predictive, adaptive, and context-aware operation. Its integration provides the necessary technological basis for the seamless convergence of \ac{TN} and \ac{NTN} in 6G, paving the way toward globally connected, intelligent, and resilient wireless systems.

\begin{figure}[t!]
	\centering
	\includegraphics[width=0.8\columnwidth]{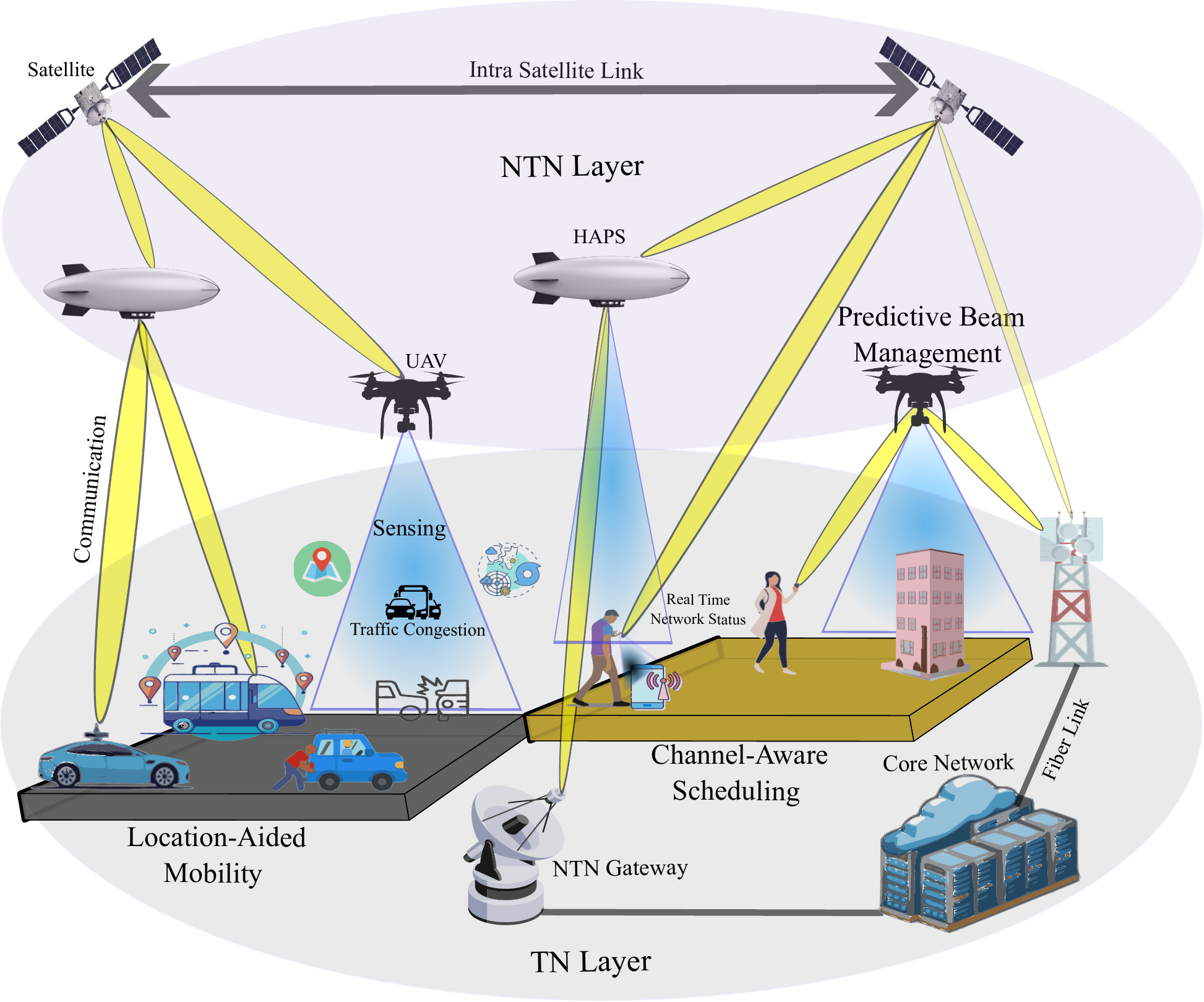}
	\caption{{ISAC-enabled NTN system architecture.}}
	\label{fige1}
\end{figure}


\begin{table*}[t]
\centering
\caption{Comparative Analysis of ISAC-Enabled NTN-Based Multiplexing Schemes}
\label{tab:comparison}

\rowcolors{2}{gray!10}{white} 

\begin{tabular}{>{\bfseries}c c c c}
\toprule
\rowcolor{gray!30}
\textbf{Parameters} & \textbf{Frequency-Division ISAC} & \textbf{Time-Division ISAC} & \textbf{Non-Orthogonal ISAC} \\
\midrule

Spectral Efficiency & Low & Low & {High} \\

Latency & Medium & High & {Low} \\

Mobility & Medium & Low & {High} \\

Complexity & Low & Low & {High} \\

Sensing Precision & Medium & Low & {High} \\

Interference & Medium & Low & {High} \\

\bottomrule
\end{tabular}
\end{table*}











\begin{figure*}[t!]
	\centering
	\includegraphics[width=0.9\linewidth]{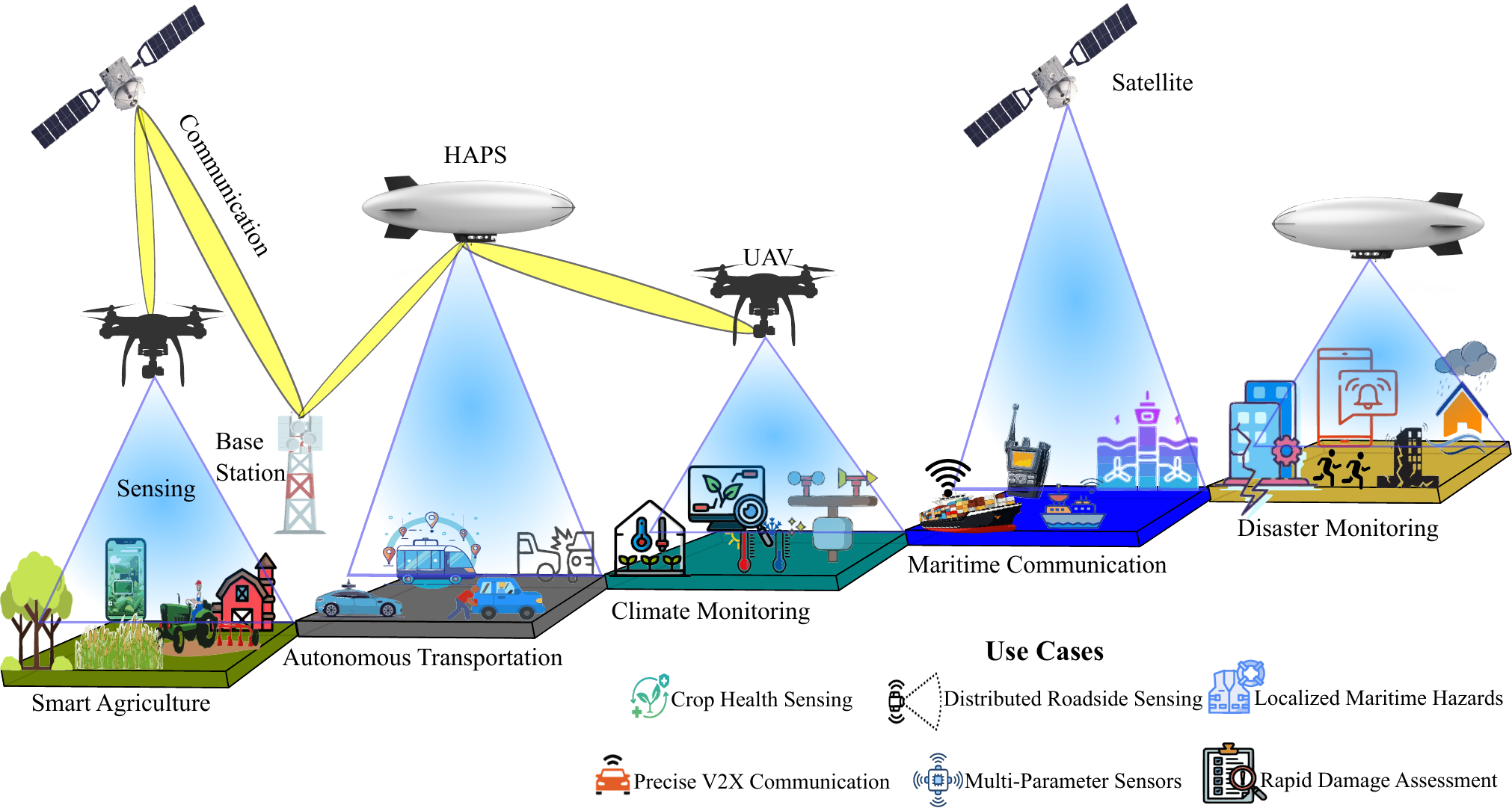}
	\caption{{ISAC-enabled NTN applications and use cases across smart agriculture, transportation, environmental monitoring, maritime services, and disaster response.}}
	\label{fige22}
\end{figure*}
\section{ISAC-Enabled NTN: System Architecture and Use Cases}

Although terrestrial \ac{ISAC} systems have demonstrated the feasibility of joint radar–communication operation, extending \ac{ISAC} to \ac{NTN} fundamentally expands the operational domain by allowing three-dimensional mobility, ultra-wide coverage, long-range propagation, and heterogeneous orbital platforms. In practice, \ac{ISAC}-enabled \ac{NTN} are naturally realized through a multi-layer \ac{NTN} architecture, where satellites, \ac{HAPS}, \ac{UAV}, and terrestrial infrastructure are jointly optimized to support both sensing and communication functionalities. This unified architecture enables the network to simultaneously provide broadband connectivity and high-resolution situational awareness across vast geographical regions. Fig.~\ref{fige1} illustrates the \ac{ISAC}-enabled \ac{NTN} architecture, where \ac{ISAC} is realized through three fundamental multiplexing paradigms:  Time-Division ISAC (TD-ISAC), Frequency-Division ISAC (FD-ISAC), and Non-Orthogonal ISAC (NO-ISAC), whose detailed comparison is provided in Table.~\ref{tab:comparison} \cite{kaushik2024integrated}. Within this architecture, satellites perform wide-area environmental sensing using a dedicated sensing waveform, while simultaneously providing communication services to \ac{UAV} and ground terminals through the same radio infrastructure. By jointly exploiting the sensing and communication capabilities, satellite platforms are able to monitor the target region and disseminate situational awareness information to \ac{UAV}, thereby enabling intelligent aerial networking and cooperative mission execution.


The space segment forms the backbone of the \ac{ISAC}-enabled \ac{NTN} architecture and consists of satellite platforms deployed across multiple orbital regimes, each offering complementary tradeoffs between coverage, latency, mobility, and sensing resolution. \ac{LEO} satellites enable low-latency \ac{ISAC} services with high mobility-induced Doppler diversity, making them particularly suitable for delay-sensitive applications such as autonomous mobility and emergency communications. \ac{MEO} satellites provide a balanced compromise between coverage footprint and sensing resolution, supporting regional-scale monitoring and navigation-assisted services. \ac{GEO} satellites offer continuous coverage in a wide-area and are mainly suitable for broadcast-oriented communication and large-scale environmental observation. From an \ac{ISAC} perspective, satellite platforms act not only as communication relays but also as large-scale sensing nodes capable of interpreting reflected radio echoes from the Earth’s surface, aerial targets, maritime vessels and atmospheric layers for localization, tracking, terrain mapping, and situational awareness.

Complementing the space layer, the aerial segment comprises near-Earth transceivers such as \ac{UAV} and \ac{HAPS}. Due to their intermediate altitude and flexible deployment, these platforms serve as adaptive \ac{ISAC} relays between the space and the ground layers. Aerial platforms enable fine-grained sensing with spatial resolution higher than that of satellites, extended coverage to remote or obstructed regions, cooperative sensing with satellites, and rapid deployment for temporary missions and emergency operations. Their mobility and on-demand reconfigurability make them particularly attractive for disaster response, maritime monitoring, border surveillance, intelligent transportation systems, and temporary hotspot provisioning.

The ground segment consists of \ac{ISAC}-enabled terrestrial \ac{BS}, edge computing nodes, and heterogeneous end-user terminals, including \ac{IoT} devices, connected vehicles, maritime sensors, and industrial monitoring systems. In addition to providing access connectivity, these nodes actively participate in the sensing process. Uplink transmissions from ground devices can be exploited for passive sensing, while dedicated probing signals enable active environmental perception. Edge computing nodes facilitate real-time data fusion, collaborative sensing, and low-latency inference, enabling mission-critical services such as cooperative perception for autonomous driving, smart port operations, precision agriculture, and large-scale infrastructure monitoring.

Beyond conventional broadband access, \ac{ISAC}-enabled \ac{NTN} are also envisioned as a global sensing and connectivity platform for massive \ac{IoT} deployments in remote and underserved regions. As \ac{IoT} sensors proliferate across mountains, forests, oceans, deserts, and rural farmlands, terrestrial infrastructure alone becomes insufficient to support the resulting connectivity and sensing demands. In this context, \ac{LEO} satellite mega-constellations, complemented by \ac{UAV} and \ac{HAPS}, enable a flexible and scalable \ac{NTN} fabric capable of simultaneously supporting massive \ac{IoT} access and wide-area environmental sensing. By leveraging \ac{ISAC}, \ac{NTN} platforms can perform remote sensing functions such as terrain mapping, vegetation monitoring, and maritime observation while maintaining reliable data links with distributed sensor nodes. Such an integrated architecture transforms \ac{NTN} into a cyber–physical sensing infrastructure that supports precision agriculture, environmental protection, and large-scale rural digitalization. In the following subsections, we explore these \ac{ISAC}-enabled \ac{NTN} use cases, also illustrated in Fig.~\ref{fige22}.

\subsection{Maritime Communication \& Surveillance}

Maritime environments represent one of the most challenging operational domains for wireless communication due to sparse terrestrial infrastructure, vast coverage areas, and highly dynamic mobility patterns. \ac{ISAC}-enabled \ac{NTN} provides broadband connectivity for ships and offshore platforms while simultaneously supporting wide-area maritime sensing. Reflected signals from vessels and sea surfaces enable real-time tracking of ship trajectories, detection of navigational hazards, and monitoring of sea-state conditions. This dual functionality improves maritime safety, supports autonomous shipping, and enables intelligent port and logistics operations.

\subsection{Disaster Monitoring \& Emergency Response}

Natural disasters such as earthquakes, floods, hurricanes, and wildfires often disrupt the terrestrial communication infrastructure, severely limiting situational awareness, coordination, and emergency logistics. In such infrastructure-compromised environments, \ac{ISAC}-enabled \ac{NTN} enable a new operational paradigm referred to as Post-Disaster \emph{Sensing-as-a-Service} (SaaS), in which sensing capabilities are dynamically provisioned as an on-demand network service alongside emergency connectivity. \ac{ISAC}-enabled \ac{NTN} therefore transforms emergency response from reactive coordination to intelligence-driven intervention. Post-disaster SaaS not only restores connectivity but also establishes a resilient, three-dimensional sensing fabric capable of delivering continuous environmental intelligence, accelerating recovery timelines, and enhancing operational safety in inaccessible or hazardous areas.



%

\subsection{Autonomous Transportation \& Air Mobility}

Future autonomous transportation systems rely on cooperative perception and high-reliability connectivity across ground vehicles, drones, and urban air mobility platforms. \ac{ISAC}-enabled \ac{NTN} extend perception and connectivity beyond terrestrial road networks, enabling seamless coverage in rural, aerial, and maritime corridors. \ac{ISAC} support cooperative object detection, trajectory prediction, and collision avoidance, forming the technological backbone of next-generation intelligent transportation systems and integrated airspace management.

\subsection{Environmental Monitoring \& Smart Agriculture}

Large-scale environmental and climate monitoring requires persistent observation over oceans, forests, glaciers, deserts, and polar regions. \ac{ISAC}-enabled \ac{NTN} transform satellites and aerial platforms into global sensing instruments capable of tracking deforestation, ice-sheet dynamics, ocean pollution, atmospheric conditions, and extreme weather events. Integration of sensing with real-time data transmission enables continuous environmental intelligence, supporting climate research, disaster prevention, and sustainable resource management. Moreover, precision agriculture and rural digitalization demand both broadband connectivity and large-area sensing capabilities. \ac{ISAC}-enabled \ac{NTN} provide coverage in remote farming regions while enabling soil monitoring, crop health assessment, irrigation optimization, and livestock tracking through aerial and satellite sensing. This integrated approach enhances food security, and accelerates digital inclusion in underserved rural communities.

\section{Standardization Activities for ISAC \& NTN}

Integration of \ac{ISAC} within \ac{NTN} introduces a paradigm shift in wireless systems. This transition necessitates a fundamental rethinking of existing standardization frameworks to jointly accommodate data transmission, environmental sensing, and mobility awareness under heterogeneous orbital and regulatory constraints. Within the \ac{6G} vision, standardization serves as a foundational enablers of interoperability, spectrum coexistence, and global adoption. 

\subsection{Current Status \& 3GPP \& ETSI Roadmaps}

Standardization has evolved along an iterative path, progressing from baseline connectivity toward integrated perception:

\begin{itemize}
    \item \textbf{Release~17 \& Release~18 (Foundational \ac{NTN}):} \ac{3GPP} Release~17 addressed the connectivity gap by integrating \ac{TN} and \ac{NTN} to provide ubiquitous coverage. Release~18 (5G-Advanced) further expanded these capabilities by incorporating initial work on \ac{AI}/\ac{ML} and wider bandwidth support to improve spectral efficiency \cite{jamshed2025tutorial}.
    
    \item \textbf{Release~19 (5G-Advanced ISAC Initiation):} \ac{3GPP} has initiated formal efforts to standardize \ac{ISAC}. In Release~19, specifically TR 22.837 in SA1, more than 30 potential use cases have been identified, including object detection, environment monitoring, and trajectory tracing \cite{3gpp22837v1940}. A critical milestone was the finalization of the \ac{ISAC} channel modeling specification in May 2025, which introduced the Extended Geometry-based Stochastic Model (E-GBSM) to ensure compatibility between communication and sensing links~\cite{3gpp38901rel19}.
    
    \item \textbf{Release~20 (5G-Advanced ISAC):} \ac{3GPP} SA2 has started architecture work on limited \ac{TN}-based \ac{ISAC} (gNB-based sensing). The study on 5G-Advanced core network support for \ac{ISAC} has been completed in TR23.700-14, while normative work on 5G-Advanced \ac{ISAC} has already begun \cite{3gpp23700-14}.
    
    \item \textbf{Release~20 (6G ISAC):} Current \ac{3GPP} Release~20  studies for \ac{6G} have accepted a total of 23 \ac{ISAC} use cases. These studies focus primarily on three application fields: \ac{UAV} tracking and management, smart transportation including autonomous driving, and industrial applications such as robotics in factories and construction environments.
    
    \item \textbf{Advanced Functionalities:} Additional capabilities are also being standardized, including target classification, accounting, sensing prediction, and the fusion of \ac{3GPP} sensing with non-\ac{3GPP} sensor systems.
    
\end{itemize}

At present, \ac{ISAC}-enabled \ac{NTN} is not within the scope of SA2 \ac{6G} study in Release~20. This implies that \ac{ISAC} in \ac{NTN} is unlikely to become a standardized \ac{6G} capability from day one. Nevertheless, this gap also highlights the need for further work to define appropriate use cases and address the remaining technical issues required to make \ac{ISAC}-enabled \ac{NTN} practically viable.

\subsection{Technical Gaps in Existing Standards}

Current standards do not yet explicitly address the unique requirements of combined \ac{ISAC} and \ac{NTN} architectures. Several important gaps remain:

\begin{itemize}

    \item \textbf{Resource Management:}  Existing frameworks lack  explicit support for \ac{ISAC}-specific control, signaling, or resource management mechanisms.
    
    \item \textbf{Waveform Constraints:} Existing multicarrier waveforms are not designed to maintain orthogonality against satellite-induced Doppler shifts, which can reach tens of kHz, and long round-trip delays, which can extend to tens of milliseconds. This leads to significant  sensing ambiguity and degraded performance.
    
    \item \textbf{Feedback Loops:} Current \ac{NTN} architectures do not provide adequate  mechanisms to convey quality indicators of detection or adapt transmission strategies based on real-time detection outcomes.
    
\end{itemize}

\subsection{Formal 3GPP Reference Summary}

Overall, current \ac{3GPP} technical reports and specifications establish the baseline for \ac{ISAC} and \ac{NTN} largely as separate technology domains. Most of definitions and use cases introduced so far remain foundational in nature.  More advanced studies and normative work specifically targeting the integration of \ac{ISAC} and \ac{NTN} are therefore expected to merge in later \ac{3GPP} releases beyond Release~21.
\subsection{ETSI Roadmap}
\begin{itemize}
    \item The \ac{ETSI} has identified key 6G \ac{ISAC} use cases, advanced channel modeling approaches, and baseline system and \ac{RAN} architectures for \ac{ISAC} in 6G. \ac{ETSI} has also highlighted important issues related to  security, privacy and trustworthiness in \ac{ISAC}-enabled systems.
    
    \item \ac{ETSI} is further working on enhancing the system and \ac{RAN} architecture for enabling \ac{ISAC} in 6G, as well as solutions for security, privacy and resilience, combined with a study of the \ac{AI}/\ac{ML} and data framework for \ac{ISAC} and the synergies of \ac{ISAC} with other key technologies, including \ac{NTN}, \ac{NFC}, \ac{RIS}, Semantic communications, and Over-the-Air Computing.
\end{itemize}


\begin{figure}
    \centering
    \includegraphics[width=0.8\columnwidth]{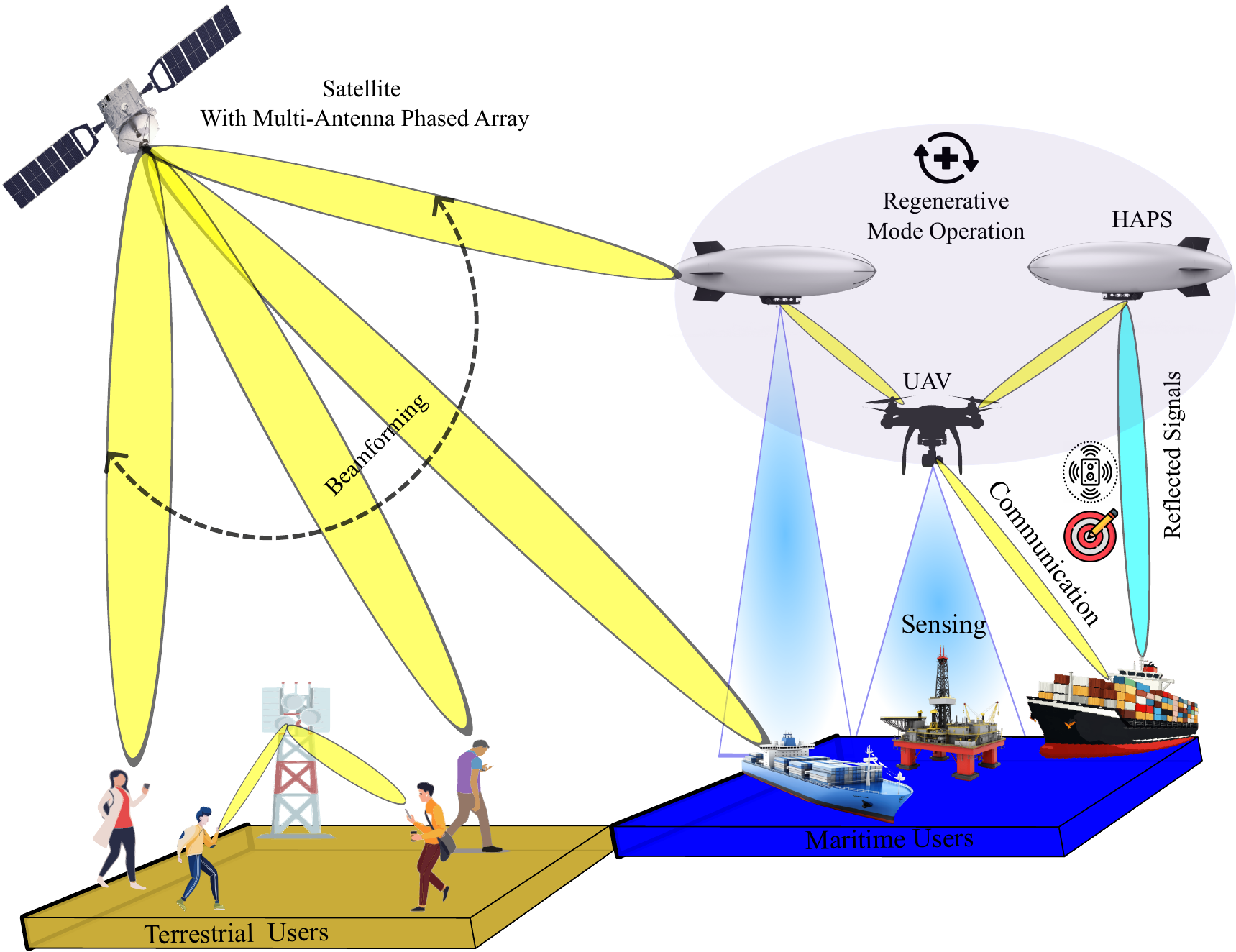}
    \caption{An illustration of ISAC assisted NTN feedback mechanism for maritime communication and situational sensing.}
    \label{fig:case}
\end{figure}

\section{Case Study: ISAC-Enabled NTN for Maritime Communication and Situational Sensing}
\label{casestudy_sec}
This section presents a representative case study that illustrates how the previously discussed concepts can be jointly utilized in an \ac{ISAC}-enabled \ac{NTN} designed to simultaneously support maritime communication and wide-area situational sensing. Based on the previous sections, several critical questions naturally arise regarding practical \ac{NTN} implementations:

\begin{itemize}
    \item Independent design of sensing and communication remains suboptimal in satellite systems. 
    \item Edge intelligence and seamless sensing feedback are essential for scalable \ac{NTN} operation.
    \item Current standards do not explicitly support \ac{ISAC}-specific control, signaling, or resource management. 
\end{itemize}

This case study delves into answering these questions, highlighting that  \ac{ISAC} operation in \ac{NTN} is fundamentally mobility-dominated, requiring new sensing and channel models tailored to highly dynamic environments. It also suggests that future \ac{6G} \ac{NTN} standards should treat \ac{ISAC} as a cross-layer system capability rather than merely a \ac{PHY} enhancement. Finally, these questions are answered at the end of this section as \textit{key takeaways}. Specifically, the maritime environment has been included to capture diverse requirements, such as sparse terrestrial requirements, large coverage requirements, dynamic targets with predictable mobility patterns, and strong sensing relevance for safety. In addition, the objective is to highlight how a unified \ac{ISAC} framework contributes to system performance compared with conventional communication-only \ac{NTN} designs.

\subsection{System Description and Assumptions} 
\textbf{Network Topology:} As illustrated in Fig. \ref{fig:case}, the considered system is assumed to consist of a satellite equipped with a multi-antenna phased array, distributed terrestrial users, maritime users, and aerial relay platforms. The \ac{ISAC}-assisted aerial stations operate in a regenerative mode, transmitting a single joint waveform that simultaneously supports  downlink communication and sensing through reflected signal processing.  

\textbf{Channel and Mobility Characteristics:} Key \ac{NTN}-specific impairments include large propagation delay, high Doppler shift due to satellite velocity, time-varying \ac{AoD}/\ac{AoA}, sea-surface reflections causing multipath propagation and clutter. 

\textbf{Joint Signal and Observation Model:} The aerial platform transmits a joint waveform $s(t)$ that carries communication symbols and sensing probes. 
The reflected signal received at the aerial station can be expressed as $r(t) = \sum_{k} \alpha_k s(t-\tau_k) e^{j2\pi f_{D,k}t}+n(t)$, where $\tau_k$ represents delay, $f_{D,k}$ represents the Doppler shift associated with relative motion, and $\alpha$ captures reflection and path-loss effects. 

\textbf{Information Duality:} Delay and Doppler simultaneously serve as communication impairments  and sensing descriptors, while beam misalignment affects both throughput and  sensing resolution. This duality enables joint estimation, eliminating the need for dedicated sensing hardware or additional signaling overhead.

\begin{figure}
    \centering
    \includegraphics[width=\columnwidth]{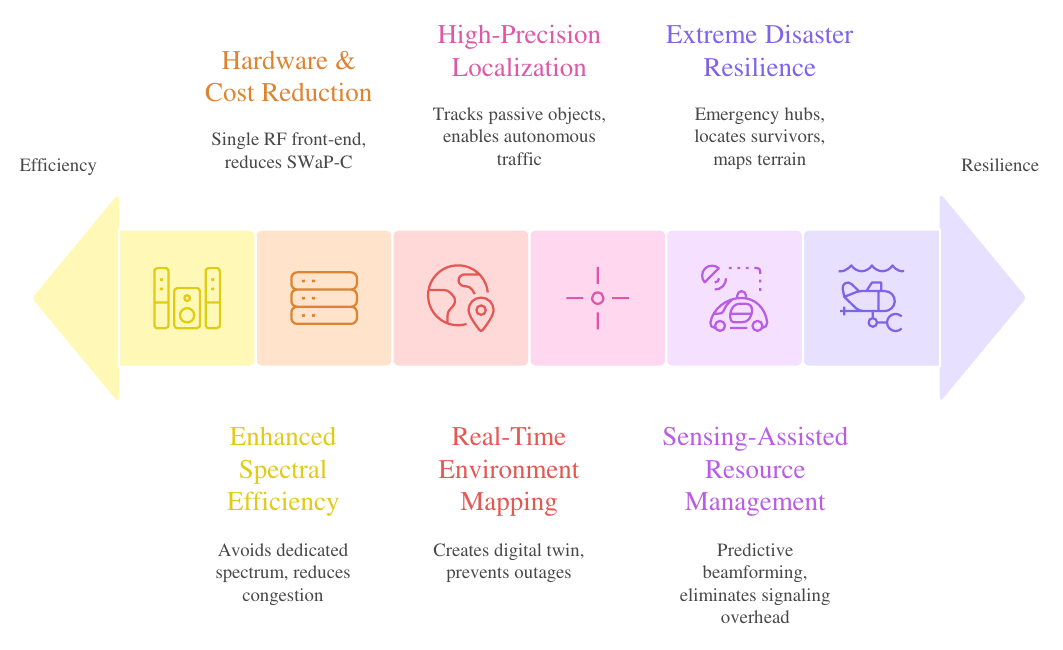}
    \caption{Key benefits of ISAC-enabled NTN, ranging from efficiency enhancement to resilience improvement.}
    \label{fig:ch}
\end{figure}

\subsection{Sensing-Assisted Communication Mechanism} 

\textbf{Target Detection and Tracking:} Reflected signals from ships and the sea surface are processed to estimate target range and velocity, angular position, and motion trajectories. Due to predictable maritime mobility patterns, Kalman or learning-based filters are effective in maintaining accurate target-state estimation. 

\textbf{Communication Adaptation:} As depicted, sensing outputs are fed into the communication control loop to predict future beam directions, pre-compensate Doppler shifts, and schedule radio resources proactively for upcoming satellite links.

\subsection{Performance Tradeoff Analysis} 
\textbf{Throughput-Sensing Tradeoff:} The presented closed-loop ISAC framework  reveals a fundamental tradeoff: increasing sensing resolution requires longer coherent processing intervals and higher pilot density, which affects the effective communication data rate. However, mechanisms such as sensing-assisted beam alignment have the potential to partially compensate for this loss by improving link quality and beam accuracy.

\textbf{Latency \& Reliability:} Compared with communication-only \ac{NTN}, \ac{ISAC}-enabled adaptation provides multi-fold benefits, including reduced outage probability, improving link reliability under high mobility, while end-to-end latency remains dominated by propagation delay but becomes more predictable. Predictability is particularly valuable for maritime safety applications.

\subsection{Core Technical Implications \& Key Takeaways} \textit{The case study reveals that sensing is not an auxiliary function but a necessary enabler for reliable communication under delayed and outdated \ac{CSI} conditions. Further, key takeaways are summarized below as:}

\begin{itemize}
    
    \item \textit{\textbf{NTN Channel Impairments Become Information Sources:} Doppler, delay, and beam drift are unavoidable in \ac{NTN} systems, hence, \ac{ISAC} converts these from errors into state variables that can be exploited for joint sensing and communication design. Unlike terrestrial \ac{ISAC}, channel estimation and detection must therefore be tightly coupled in \ac{NTN} environments.}
    
    \item \textit{\textbf{ISAC Enables Predictive Rather Than Reactive Communication:} In \ac{NTN} systems, feedback delay often exceeds channel coherence time owing to the large Doppler and long propagation delays. As a result, conventional \ac{CSI}-based reactive adaptation becomes ineffective, whereas  \ac{ISAC} enables predictive control-based environmental state estimation.}
    
    \item \textit{\textbf{Unified Waveform Reduces Payload and Signaling Overhead:} Separate sensing payloads increase satellite mass, power consumption, and hardware complexity. Therefore, a unified \ac{ISAC} waveform provides a hardware-efficient and spectrum-efficient solution that is particularly attractive for satellite platforms.}
    
\end{itemize}

\section{Challenges and Future Considerations} 
The integration of \ac{ISAC} into \ac{NTN}  fundamentally transforms the design assumptions of both sensing and communication systems. Beyond terrestrial \ac{ISAC}, \ac{NTN}-based \ac{ISAC} brings additional challenges in the form of high mobility, large-scale propagation, and strict payload constraints, necessitating a rethinking of signal models, architectures, and control mechanisms. At the same time, it brings recalls for spectrum utilization, hardware considerations, real-time service delivery. As depicted in Fig. \ref{fig:ch}, this section presents critical challenges and future considerations in this direction.
\vspace{-0.15cm}
\subsection{ISAC Waveform Design under Extreme Doppler and Delay Spread} 
In \ac{NTN} environments, relative satellite–air stations–user motion induces excessive Doppler shifts ranging from several kHz to tens of kHz, and the round-trip delays can exceed tens of milliseconds \cite{sat}. This, in turn, effects orthogonality of multicarrier waveforms and channel reciprocity assumptions. Further, \ac{ISAC} waveforms optimized for terrestrial environments cannot be directly applied to \ac{NTN} scenarios owing to the severe sensing ambiguity degradation, and reduced detection probability.

\textbf{Potential Solutions:} Potential solutions include the careful design of delay–Doppler-native \ac{ISAC} waveforms. Recent efforts like adaptive waveform reconfiguration and joint waveform–receiver co-design open futuristic research directions.  
\vspace{-0.15cm}
\subsection{Channel Modeling \& Sensing–Communication Coupling} Another major challenge arises in the form of accurate channel models as \ac{NTN} channels exhibit a rapid time variation, and \ac{LoS} dominance with time-varying angles, affecting metrics like localization precision \cite{ang1}. Thereby classical \ac{ISAC} assumptions are no longer applicable to these frameworks.

\textbf{Potential Solutions:} Among probable solutions lie a unified channel–sensing modeling strategy capturing geometry, mobility, and reflection dynamics. Besides, joint estimation mechanisms capable of treating communication channel estimation and sensing parameter as a single inverse problem.
\vspace{-0.15cm}
\subsection{Beamforming On-Board Processing Constraints} In satellite systems, beams serve dual purposes while narrow beams improve sensing angular resolution, wider beams improve communication coverage. Moreover, \ac{NTN} payloads face strict limits on processing capability, energy availability, and thermal dissipation. Thereby, transmitting raw sensing data to ground stations may cause excessive backhaul load and increased end-to-end latency.

\textbf{Potential Solutions:} Modern approaches including multi-objective beamforming capable of jointly optimizing features like angular resolution and coverage probability, sensing assisted resource management. Moreover, approaches like compressed sensing and feature extraction before transmission proven to be effective under these scenarios.
\vspace{-0.15cm}
\subsection{Control Signaling and Interference Management} Current \ac{NTN} architectures lack mechanisms to convey signal sensing quality indicators, adapt transmission strategies based on sensing outcomes, and hence, \ac{ISAC} often operates in an open-loop manner, limiting performance gains. Further, \ac{ISAC} transmissions in \ac{NTN} prone to terrestrial interference, particularly vulnerable to interference due to their reliance on weak reflections \cite{nt12}.

\textbf{Potential Solutions:} Necessary transformations including the definition of sensing-aware control channels, feedback mechanisms for sensing are required to be adopted to mitigate controlling challenges. Furthermore, interference management can be mitigated through coordination mechanisms between \ac{NTN} and \ac{TN} systems, necessitating regulatory-compliant sensing-aware transmission policies.

\section{Conclusion}
The integration of \ac{ISAC} into \ac{NTN} marks a fundamental transition from communication-centric network design toward perception-driven network operation. Instead of optimizing solely for throughput and latency, future \ac{NTN} systems will jointly consider environmental awareness, mobility prediction, and sensing accuracy as intrinsic performance metrics. This transformation establishes \ac{ISAC} as a foundational capability for \ac{NTN} rather than a supplementary feature. In this context, \ac{ISAC}-enabled \ac{NTN} are envisioned as a cornerstone technology for realizing ubiquitous, resilient, and intelligent \ac{6G} connectivity. However, standardization gaps still remain, which may limit the potential of \ac{NTN} and \ac{ISAC}. To fully unlock the potential of \ac{6G}, \ac{ISAC} must therefore be systematically incorporated into integrated \ac{TN} and \ac{NTN} systems.

\bibliographystyle{IEEEtran}

\bibliography{IEEEabrv,BibRef}

\begin{thebibliography}{10}
\providecommand{\url}[1]{#1}
\csname url@samestyle\endcsname
\providecommand{\newblock}{\relax}
\providecommand{\bibinfo}[2]{#2}
\providecommand{\BIBentrySTDinterwordspacing}{\spaceskip=0pt\relax}
\providecommand{\BIBentryALTinterwordstretchfactor}{4}
\providecommand{\BIBentryALTinterwordspacing}{\spaceskip=\fontdimen2\font plus
\BIBentryALTinterwordstretchfactor\fontdimen3\font minus \fontdimen4\font\relax}
\providecommand{\BIBforeignlanguage}[2]{{%
\expandafter\ifx\csname l@#1\endcsname\relax
\typeout{** WARNING: IEEEtran.bst: No hyphenation pattern has been}%
\typeout{** loaded for the language `#1'. Using the pattern for}%
\typeout{** the default language instead.}%
\else
\language=\csname l@#1\endcsname
\fi
#2}}
\providecommand{\BIBdecl}{\relax}
\BIBdecl

\bibitem{jamshed2025tutorial}
M.~A. Jamshed, A.~Kaushik, S.~Manzoor, M.~Z. Shakir, J.~Seong, M.~Toka, W.~Shin, and M.~Schellmann, ``A tutorial on non-terrestrial networks: Towards global and ubiquitous {6G} connectivity,'' \emph{Foundations and Trends in Networking}, vol.~14, no.~3, pp. 160--253, 2025.

\bibitem{11372048}
M.~Ali~Jamshed, M.~Ahmed~Mohsin, H.~Zhang, B.~Haq, A.~Kaushik, B.~Di, and W.~Jiang, ``Reconfigurable holographic surfaces and near field communication for non-terrestrial networks: Potential and challenges,'' \emph{IEEE Wireless Communications}, pp. 1--8, 2026.

\bibitem{kaushik2024integrated}
A.~Kaushik, R.~Singh, M.~Li, H.~Luo, S.~Dayarathna, R.~Senanayake, X.~An, R.~A. Stirling-Gallacher, W.~Shin, and M.~Di~Renzo, ``Integrated sensing and communications for {IoT}: Synergies with key {6G} technology enablers,'' \emph{IEEE Internet of Things Magazine}, vol.~7, no.~5, pp. 136--143, 2024.

\bibitem{meng2023uav}
K.~Meng, Q.~Wu, J.~Xu, W.~Chen, Z.~Feng, R.~Schober, and A.~L. Swindlehurst, ``{UAV}-enabled integrated sensing and communication: Opportunities and challenges,'' \emph{IEEE Wireless Communications}, vol.~31, no.~2, pp. 97--104, 2023.

\bibitem{kanani2025haps}
P.~Kanani, M.~J. Omidi, M.~Modarres-Hashemi, and H.~Yanikomeroglu, ``{HAPS-ISAC} for {6G}: Architecture, design trade-offs, and a practical roadmap,'' \emph{arXiv preprint arXiv:2510.23147}, 2025.

\bibitem{yin2024integrated}
L.~Yin, Z.~Liu, M.~B. Shankar, M.~Alaee-Kerahroodi, and B.~Clerckx, ``Integrated sensing and communications enabled low earth orbit satellite systems,'' \emph{IEEE Network}, vol.~38, no.~6, pp. 252--258, 2024.

\bibitem{11098638}
Y.~Song, Y.~Zeng, Y.~Yang, Z.~Ren, G.~Cheng, X.~Xu, J.~Xu, S.~Jin, and R.~Zhang, ``An overview of cellular {ISAC} for low-altitude {UAV}: New opportunities and challenges,'' \emph{IEEE Communications Magazine}, vol.~63, no.~12, pp. 88--95, 2025.

\bibitem{11017717}
D.~He, W.~Yuan, J.~Wu, and R.~Liu, ``Ubiquitous {UAV} communication enabled low-altitude economy: Applications, techniques, and {3GPP}’s efforts,'' \emph{IEEE Network}, vol.~40, no.~1, pp. 115--122, 2026.

\bibitem{10463684}
M.~A. Jamshed, A.~Kaushik, M.~Toka, W.~Shin, M.~Z. Shakir, S.~P. Dash, and D.~Dardari, ``Synergizing airborne non-terrestrial networks and reconfigurable intelligent surfaces-aided {6G} {IoT},'' \emph{IEEE Internet of Things Magazine}, vol.~7, no.~2, pp. 46--52, 2024.

\bibitem{3gpp22837v1940}
\BIBentryALTinterwordspacing
{3GPP}, ``{Study on Integrated Sensing and Communication (Release 19)},'' 3GPP Technical Report (TR) 22.837, Jun. 2024, release 19. [Online]. Available: \url{https://www.3gpp.org/dynareport/22837.htm}
\BIBentrySTDinterwordspacing

\bibitem{3gpp38901rel19}
------, ``{Study on ISAC Channel Model (Release 19)},'' {3rd Generation Partnership Project (3GPP)}, France, Technical Report TR 38.901, May 2025, version 19.x.x.

\bibitem{3gpp23700-14}
\BIBentryALTinterwordspacing
------, ``Study on stage 2 for integrated sensing and communication (isac),'' 3rd Generation Partnership Project (3GPP), Technical Report TR 23.700-14, 2025. [Online]. Available: \url{https://www.3gpp.org/dynareport/23700-14.htm}
\BIBentrySTDinterwordspacing

\bibitem{sat}
O.~Kodheli, E.~Lagunas, N.~Maturo, S.~K. Sharma, B.~Shankar, J.~F.~M. Montoya, J.~C.~M. Duncan, D.~Spano, S.~Chatzinotas, S.~Kisseleff \emph{et~al.}, ``Satellite communications in the new space era: A survey and future challenges,'' \emph{IEEE Communications Surveys \& Tutorials}, vol.~23, no.~1, pp. 70--109, 2020.

\bibitem{ang1}
A.~Shahmansoori, G.~E. Garcia, G.~Destino, G.~Seco-Granados, and H.~Wymeersch, ``Position and orientation estimation through millimeter-wave {MIMO} in {5G} systems,'' \emph{IEEE Transactions on Wireless Communications}, vol.~17, no.~3, pp. 1822--1835, 2017.

\bibitem{nt12}
F.~Rinaldi, H.-L. Maattanen, J.~Torsner, S.~Pizzi, S.~Andreev, A.~Iera, Y.~Koucheryavy, and G.~Araniti, ``Non-terrestrial networks in {5G} \& beyond: A survey,'' \emph{IEEE Access}, vol.~8, pp. 165\,178--165\,200, 2020.

\end{thebibliography}

\end{document}